\documentclass[preprint,flushrt]{aastex}
\usepackage{graphicx}
\newcommand{\eqb}{\begin{eqnarray}}
\newcommand{\eqe}{\end{eqnarray}}
\newcommand{\diff}{{\rm d}}
\newcommand{\rlight}{r_{\rm L}}
\newcommand{\blight}{B_{\rm L}}
\newcommand{\nlight}{n_{\rm L}}
\newcommand{\slight}{\sigma_{\rm L}}
\newcommand{\olight}{\omega_{\rm L}}
\newcommand{\glight}{\gamma_{\rm L}}
\newcommand{\unitphi}{{\bf \hat\varphi}}
\newcommand{\unittheta}{{\bf \hat\theta}}
\newcommand{\unitr}{{\bf \hat r}}
\newcommand{\ncold}{n'_{\rm c}}
\newcommand{\nhot}{n'_{\rm h}}
\newcommand{\conduct}{\sigma_{\rm c}}
\newcommand{\vwave}{v_{\rm w}}
\newcommand{\rslow}{R}
\newcommand{\eps}{\varepsilon}

\newcommand{\rnonrel}{r_2}
\slugcomment{submitted to ApJ}

\begin{document}
\title{Reconnection in a striped pulsar wind}
\author{Y. Lyubarsky} 
\affil{Department of Physics, Ben-Gurion University, P.O. Box 653, Beer Sheva 84105, Israel}
\email{lyub@bgumail.bgu.ac.il}
\and
\author{J. G. Kirk}
\affil{Max-Planck-Institut f\"ur Kernphysik,
Postfach 10 39 80, 69029 Heidelberg, Germany}
\email{John.Kirk@mpi-hd.mpg.de}
\begin{abstract}
It is generally thought that most of the spin-down power of a pulsar 
is carried away in an MHD wind dominated by Poynting flux. 
In the case of an oblique rotator, 
a significant part of this energy can be considered 
to be in a low-frequency wave, consisting of stripes of 
toroidal magnetic field of alternating polarity, propagating in a 
region around the equatorial plane. Magnetic reconnection 
in such a structure has been proposed as a mechanism for 
transforming the Poynting flux
into particle energy in the pulsar wind. We have re-examined this process
and conclude that the wind accelerates significantly in the course of
reconnection. This dilates the timescale over which the reconnection 
process operates, so that the wind requires a much larger distance 
than was previously thought in order to convert the Poynting flux to 
particle flux. In the case of the Crab, the 
wind is still Poynting-dominated
at the radius at which a standing shock is inferred from observation.
An estimate of the radius of the termination shock for other 
pulsars implies that all except the milli-second pulsars 
have Poynting-flux dominated winds all the way out to the shock front.  
\end{abstract}

\section{Introduction}
\label{introduction}
The diffuse synchrotron radiation from the Crab Nebula has now been observed in great detail in 
several wavebands, e.g., \citet{hesteretal95}. 
Although by far the best observed example, the Crab is 
just one of several pulsars surrounded by such a nebula
\citep{arons96}, whose ultimate source of energy is 
almost certainly the rotational kinetic energy of the central neutron star
\citep{pacini67}.
However, despite intensive efforts, it is still not known how this energy is 
transferred from the neutron star to the radiating, relativistic electrons. 

It is widely accepted that pulsars emit an 
electron-positron plasma, which carries away energy in the 
form of an 
ultra-relativistic
magnetized wind, together with large amplitude waves \citep{reesgunn74}.
At a shock front, located, in the case of the Crab Nebula, some 
$10^{17}\,$cm from the neutron star, this energy is released into 
the relativistic electrons responsible for the observed radiation. 
The most serious problem with this picture is that 
close to the strongly magnetized neutron star the 
energy must be carried mostly by 
electromagnetic fields as Poynting flux \citep{michel82}. But, in 
order to produce the 
radiating electrons, the energy flux at the shock front must be 
carried mainly by the particles
\citep{reesgunn74,kennelcoroniti84,emmeringchevalier87}.
As has been pointed out by several authors --- and emphasized recently by 
\citet{chiuehetal98} and \citet{bogovalovtsinganos99} --- in an 
ideal, ultra-relativistic MHD wind, there is no 
plausible way of converting the Poynting flux into particle energy flux.
Several effects have been investigated in attempts 
to overcome this difficulty, ranging from rapid expansion in a magnetic nozzle
\citep{chiuehetal98} to non-ideal MHD effects in a two-fluid 
(electron and positron) plasma \citep{mestelshibata94,melatosmelrose96}. 
It has also been 
suggested that a global non-axisymmetric instability of the nebular plasma 
alleviates the problem by enabling 
the conversion of magnetic to particle 
energy \citep{lyubarsky92,begelman98}. However, perhaps the 
most promising explanation to date has been that of reconnection in a 
striped pulsar wind, as advanced by \citet{michel82}, \citet{coroniti90} and 
\citet{michel94}. Here, we reanalyze this process. Both Coroniti and
Michel effectively assumed the wind maintained constant 
speed during reconnection. 
We find that the hot plasma in the current sheets 
where reconnection takes place performs work on the wind, accelerating it 
substantially. As a result,  
the reconnection rate is 
much slower than hitherto claimed. In the case of the Crab Nebula, 
we conclude that reconnection is not 
capable of converting a significant fraction of the energy flux before the 
wind reaches the position at which the shock front is encountered. 

The conversion of Poynting flux to kinetic energy is also a central problem
in a certain class of models of gamma-ray bursts. It has been proposed that
the emission of a large amplitude electromagnetic wave by 
a black hole or neutron star(s) underlies the burst phenomenon, that
the gamma-rays are generated by dissipative processes in the wave, and 
that the 
X-ray and lower frequency emission is produced at an outer 
shock front
\citep{usov92,usov94,blackmanyi98}. 
The calculations we present can be rescaled to this situation and may offer 
a mechanism for accelerating the wind to the very high Lorentz factors
required before dissipation sets in.

The paper is organized as follows: in Section~\ref{coroniti} we describe 
the problem as formulated by \citet{coroniti90} and \citet{michel94},
present estimates of the effect of reconnection and 
explain in physical terms the reasons for 
our new result. The perturbation method we employ is a short 
wavelength approximation, which 
uses as a small parameter the 
ratio of the radius of the 
light cylinder to the actual radius. This is  
presented in outline in Section~\ref{perturbation},
relegating most of the algebra to the Appendix, but pointing out the 
differences between our method and that of \citet{coroniti90}.
An analytic asymptotic solution 
to the system is given in Section~\ref{results}, followed by
results of 
a numerical integration of the 
equations and a discussion of the limitations of our approach.
Finally, in Section~\ref{conclusions}, 
we summarize our conclusions and their implications, especially
for our understanding of the Crab Nebula.

\section{The striped wind}
\label{coroniti}
 
Despite the presence of a plasma, the total power 
lost by a rotating magnetized neutron star may be
estimated using the formula for a 
magnetic dipole rotating in vacuum \cite{gunnostriker69}, \cite{michel82}. 
However, the presence of plasma changes the physical picture
significantly: for example, even 
an axisymmetric rotator surrounded by plasma 
loses energy by driving a plasma wind, 
in contrast to an aligned 
magnetic dipole in vacuum. 
Furthermore, although a strong vacuum
electromagnetic wave readily loses energy to particles
\citep{ostrikergunn69, asseoetal78, melatosmelrose96},
the presence of 
plasma is likely to prevent the formation of such waves and restrict 
dissipation to shock fronts or regions in which the magnetic field 
undergoes reconnection. 

In the case of an oblique rotator, the energy lost in the wind 
can be regarded as shared
between an axisymmetric component of the Poynting flux 
and one due to MHD waves, the ratio
being determined by the angle between the magnetic and rotational
axes.
\citet{michel71} pointed out that such waves, which have a phase speed less
than that of light, should evolve into 
regions of cold magnetically dominated
plasma separated by very narrow, hot, current sheets. 
The formation of this pattern may be imagined as follows: 
in the axisymmetric
case, a current sheet separates 
the two hemispheres with opposite polarity beyond the light
cylinder.
As the obliquity increases, 
this sheet begins to oscillate about the 
equatorial plane because the
field line at a given radius alternates in direction with the frequency
of rotation, being connected to a different magnetic pole every
half-period. Such a picture
is observed in Solar Wind. In a quasi-radial flow, the amplitude of these
oscillations grows linearly with radius and at large distances
one can locally imagine quasi-spherical current sheets following
each another and separating the stripes of magnetized plasma 
with opposite polarity.
\citet{coroniti90} called this picture a {\em striped wind}. Recently 
\citet{bogovalov99} has found an exact solution for the oblique split
monopole case which has precisely this structure.

It was noticed by \cite{usov75} and 
\citet{michel82} that these waves must decay
at large distances, since the current required to sustain them falls off as
$r^{-1}$. This is slower than the fall-off in the 
available number of charge carriers, which
goes as $r^{-2}$.
\citet{coroniti90}
considered the reconnection process in a striped wind 
and came to the same conclusion. In the case of a highly oblique rotator, 
both Coroniti and Michel agreed that the MHD wind 
of the Crab pulsar could be transformed by this mechanism
from one dominated by Poynting flux to one dominated by particle
kinetic energy flux well within the radius of the termination shock.

The pulsar wind beyond the light cylinder is quasi-spherical
\citep{chiuehetal98, bogovalovtsinganos99}, 
and, as the magnetic field is predominantly toroidal, it scales
roughly as
\eqb
B&=&\blight\frac{\rlight}{r},
\label{magscaling}
\eqe
where $\rlight=c/\Omega=cP/2\pi$ is the light-cylinder radius, $\Omega$ the 
angular velocity of the neutron star $P$ its period of rotation and $\blight$
the magnetic field strength at $\rlight$. 
Near the
light cylinder, the poloidal and toroidal components are comparable, but
within it the poloidal field is nearly dipolar, so that
one can estimate
\eqb
\blight&=&\frac{\mu}{\rlight^3}\,\approx\,9\frac{\mu_{30}}{P^3}{\rm G},
\label{bestimate}
\eqe
where $\mu=10^{30}\mu_{30}\,$G$\cdot$cm$^3$ is the magnetic moment of
the star and $P$ is given in seconds.
The density $n$ in the quasi-spherical flow, 
measured in the rest frame of the star, decreases approximately as
\eqb
n&=&\nlight\left(\frac{\rlight}{r}\right)^2,
\label{denscaling}
\eqe
since in a steady, 
spherical, relativistic flow $r^2 n v$ is constant and 
the speed $v$ is close to $c$. The
density is conveniently normalized by the Goldreich-Julian
charge density $\rho_{\rm GJ}$:
\eqb
\nlight&=&\kappa \rho_{\rm GJ}/e
\nonumber\\
&=&\frac{\kappa \blight\Omega}{2\pi ec}
\label{gjdensity}
\eqe
where $\kappa$ is the {\em multiplicity coefficient}. 
This quantity is rather uncertain
but generally expected to be large: $\kappa\sim 10^3-10^4$
\citep{arons83}. 

The Poynting flux may be estimated as
\eqb
W&=&\frac{c\blight^2}{4\pi}\left(\frac{\rlight}{r}\right)^2
\label{poynting}
\eqe
and the ratio of the Poynting flux to the kinetic energy flux,
called the magnetization parameter $\sigma$,
is
\eqb
\sigma&=&\frac W{mc^3\gamma n}
\enspace.
\label{sigma}
\eqe
At the light cylinder this quantity takes on the value
\eqb
\slight&=& {\olight \over 2\glight\kappa\Omega}
\label{sigmal}\\
&=&1\cdot3\times10^{7} {\mu_{30}\over \glight\kappa P^2}
\eqe
where $\olight=e\blight/mc$ is 
the gyrofrequency at the light cylinder.
The ratio $\olight/\Omega$ is large for all pulsars. 
For example, in the case of the Crab, where
$\mu_{30}\approx5$, we have 
$\olight/\Omega\approx 10^{11}$.

The speed of a fast magnetosonic wave propagating perpendicular to the
magnetic field in a magnetically dominated plasma corresponds to a Lorentz
factor $\gamma_{\rm fms}=\sqrt{\sigma}$ [e.g., \citet{kirkduffy99}]. 
In pulsar models, plasma is ejected at Lorentz factors of 
$10^2$ to $10^3$, which is higher than $\gamma_{\rm mfs}$  
for most pulsars. In the case of the Crab, however, these values correspond
to  a trans-alfv\'enic speed 
($\glight\sim\sqrt{\slight}$) at the 
light cylinder. At some point beyond the light cylinder, but 
before the onset of dissipation, 
we expect the wind to establish
a supersonic ideal MHD flow, in which the value of $\sigma$ and $\gamma$
 are constant. In the following, these values are indicated by the subscript \lq L\rq, 
even though, strictly speaking, 
they may not be achieved at the light cylinder itself. 
Using the above estimates, we then find that, 
for the Crab,
$\kappa=10^4$ corresponds to $\slight\approx3\times10^4$ and 
$\glight\approx200$.
At the light cylinder, almost all the 
energy is carried by the electromagnetic field.
If this energy were completely transferred from Poynting flux into plasma kinetic
energy, the Lorentz factor would attain the value
\eqb
\gamma_{\rm max}&=&\glight\slight
\nonumber\\
&=&\frac{\olight}{2\kappa\Omega}.
\label{gammamax}
\eqe
However, a cold, radial MHD wind does not accelerate, because the outwardly directed 
pressure gradient 
arising from the gradient of the magnetic field is exactly balanced by the inwardly directed 
tension force exerted by the curved toroidal field lines. In the absence of dissipation, 
the energy flux remains locked in the field as Poynting flux. 

\begin{figure}
\includegraphics[bb=138 393 526 630,width=\textwidth]{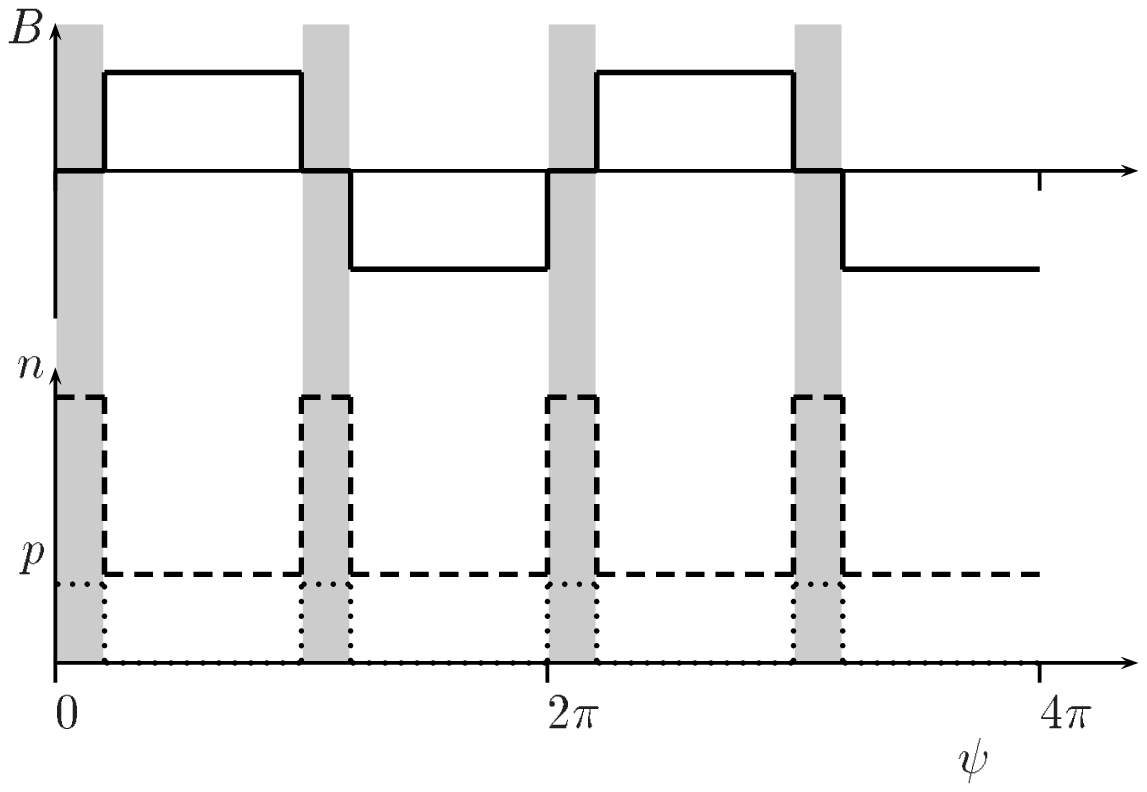}
\caption{\protect\label{sketch}
Idealized picture of a pulsar wind driven by a 
rotating, perpendicular, split-monopole magnetic field, 
showing the magnetic field, $B$ (solid line), particle density $n$ 
(dashed line) and 
plasma pressure $p$ (dotted line) as functions of the phase variable
$\phi=r/\rlight$, with origin at an arbitrary position in the 
outer part of the wind ($r\gg\rlight$).
The magnetic field and density are constant between the current sheets,
which are shown shaded.
Within the sheets, we assume zero magnetic field and constant density.} 
\end{figure}

The idealized radial structure of the wind and embedded current sheets corresponding to 
the field of a perpendicular split monopole is shown in Fig.~\ref{sketch}.
In the proper plasma frame the current density $I'$ 
(current per unit length in the toroidal direction)
is, according to Amp\`ere's law,
\eqb
I'&=&\frac{cB'}{2\pi}.
\label{current}
\eqe
Here and in the following we use primed values to indicate quantities in the proper plasma
frame, e.g., 
$B'=B/\gamma$, $n'=n/\gamma$ etc.  
\citet{usov75} and \citet{michel82} noticed that 
the current in Eq.~(\ref{current}) cannot be maintained to arbitrarily 
large radius: since the proper current density cannot
exceed $en'c$ and the width of the sheet is less than half a wavelength,
$\pi \rlight$, one can easily see 
from Eqs.(\ref{magscaling}), (\ref{denscaling}) and (\ref{gjdensity}) 
that the striped wind cannot exist beyond the radius
\eqb
r_{\rm max}&=&\pi\kappa\gamma \rlight
\label{rmax}
\eqe
[see also \citep{melatosmelrose96}].
When the velocity of the current carriers  
approaches the speed of light, an anomalous
resistance arises and the alternating magnetic field 
dissipates by heating the plasma. \citet{coroniti90}
considered the dissipation as reconnection through the current sheet,
whose minimum thickness he took equal to the Larmor radius 
which a thermal particle of 
the hot plasma in the sheet would have, if it entered the cold magnetized part of the wind. 
In terms of the fraction $\Delta$ of 
a wavelength $2\pi\rlight$ 
occupied by the two current sheets, this condition reads
\eqb
\Delta&>& {T'\over \pi \gamma\rlight eB'}
\label{larmor}
\eqe
where $T'$ is the characteristic ``temperature'' of the particles in the sheet,
in energy units.
Taking into account that the sheet is in pressure equilibrium,
\eqb
n'T'&=&{B'^2\over 8\pi}
\label{pressure}
\eqe
one can easily see that the current density 
$j'<en'c$, so that Coroniti's condition is, 
to within a factor of the order of unity, the same as that of
\citet{michel94}.

If, initially, a fraction $\alpha$ of the plasma is concentrated in
the sheets, dissipation begins at a radius $r_1=\alpha r_{\rm max}$.
For $r<r_1$ the sheet width exceeds the limiting value of Eq.~(\ref{larmor}),
the current carriers remain non-relativistic, the conductivity is very high 
and reconnection does not proceed. 
For $r>r_1$, however, the current carriers become relativistic,
an anomalous resistivity arises, and reconnection brings energy and plasma into
the sheet until, eventually, all the magnetic flux is destroyed at
the radius $r=r_{\rm max}$.

Both Michel and Coroniti estimated the radius $r_{\rm max}$ 
by substituting
into Eq.(\ref{rmax}) the initial Lorentz factor $\glight$. The corresponding
value is large compared to the light cylinder radius. For the Crab, it is 
of the order of 
$10^7\rlight$, which is, nevertheless, still well within the radius of the 
standing shock $r_{\rm s}\approx 10^9\rlight$.
However, as we show below, the flow accelerates as reconnection proceeds. 
Basically, this is because the hot plasma continues to exert an outwardly 
directed pressure gradient, but the compensating inwardly directed tension force is 
absent. Equivalently, it is clear that
the hot plasma performs work during the radial expansion,
and this appears as an acceleration of the wind. 
In an accelerating wind, most of the energy is
released when the Lorentz factor of the flow is roughly at its 
maximum value, given by Eq.~(\ref{gammamax}). The corresponding radius is
\eqb
r_{\rm max}&=&
\pi\kappa\gamma_{\rm max}\rlight
\nonumber\\
&=&\frac{\pi\olight}{2\Omega}\rlight
\label{correctrmax}
\\
&=&2\cdot 10^{17}\frac{\mu_{30}}P\,{\rm cm}
\nonumber
\eqe
In the case of the Crab, $r_{\rm max}\approx 10^{11}\rlight$, which significantly exceeds the radius of the 
standing shock,
so that only a small fraction of the 
wave energy is converted into particle energy before the plasma 
arrives at this shock front.

\section{Equations}
\label{perturbation}
\subsection{Local structure of the wave}
The striped wind described above can be regarded as an MHD wind 
containing an entropy wave which moves together with the plasma. 
At large distances the wave consists locally of spherical current sheets 
separating cold, magnetized stripes of plasma with opposing polarity 
(see Fig.~\ref{sketch}). 
To find the evolution of this wave, we use a two-timescale perturbation 
approach, assuming that the timescale on which it evolves i.e., the timescale
on which the reconnection sheet grows, is much longer than the 
period $P$ of the wave. Details of this method are presented in the Appendix.
Here we restrict ourselves to the most important steps in the argument.

In the cold magnetized part of the wind the magnetic field $B$, which is 
toroidal, is constant on the fast timescale and the plasma
pressure vanishes. The
plasma density $\ncold$ ($n_{\rm c}$ in the lab.\ frame) 
is also constant on the fast timescale.
In the hot sheet, the magnetic field vanishes, the pressure $p'$ is constant
on the fast timescale, as is the plasma density $\nhot$, which, in
contrast to previous treatments, is not constrained to equal $\ncold$.  
In the plasma frame the entire pattern is at rest, but
the plasma speed $v$, and the corresponding Lorentz factor $\gamma$  
change on the slow timescale as the wave evolves. 
In the cold part of the wind, the magnetic field is frozen into the plasma, so
that in spherical polar coordinates we can write for the slow 
evolution in radius
\eqb
{\diff\over\diff r}\left({B\over r n_{\rm c}}\right)&=&0.
\label{textfreeze1}
\eqe
[see Eq.~(\ref{freeze})].
Although the method is strictly valid only for $r\gg\rlight$, 
we normalize the quantities in the wind to their value at $r=\rlight$. Using 
Eq.~(\ref{gjdensity}) and interpreting  
the multiplicity factor as referring to the density of 
electron/positron pairs outside the 
current sheets, this yields
\eqb
{B'\over r \ncold}&=&{2\pi e\over \kappa}.
\label{textfreeze}
\eqe
Following \citet{coroniti90}, we assume that reconnection
keeps the sheet width equal to the limiting value given in
Eq.~(\ref{larmor}) and (\ref{coroniticond}). 
The condition that the cold and hot parts of the
flow are in pressure equilibrium (\ref{pressure}) leads, 
together with Eq.~(\ref{textfreeze}), to the relation
\eqb
\Delta&=&{r Z c\over 4\pi\gamma v \rlight\kappa}
\label{textcoroniticond}
\eqe
where a useful variable
\eqb
Z&=&{\ncold \over \nhot}\,=\,{n_{\rm c}\over n_{\rm h}}
\label{defZ}
\eqe
has been introduced.
In his treatment, \citet{coroniti90}  set
$\nhot=\ncold$ to write his equation (16) 
--- the counterpart 
of our Eq.~(\ref{textcoroniticond}).

The remaining equations needed to describe the wave evolution are those of
conservation of particle number and energy (equivalent to the equation of
motion in the relativistic formulation) and the entropy equation. 
To zeroth order in $\rlight/r$
these confirm that the plasma speed equals the pattern speed,
the hot and cold parts are in pressure equilibrium and 
the configuration shown in Fig.~\ref{sketch} and described above 
is stationary on the fast timescale. 

\subsection{The continuity equation}
The slow evolution of the system is obtained by 
averaging over a wave-period, which we denote by angle brackets: 
$\left<\dots\right>$. Conservation of
particle number (the continuity equation) gives
\eqb
\left<nvr^2\right>&=&{\rm constant}
\eqe
This equation is exact. In the short wavelength approximation, one can 
replace the time average by a spatial one, because all slowly varying
parameters are constants on the scale of one wavelength. This immediately
yields 
\eqb
vr^2 n_{\rm c}\left[{\Delta\over Z}+\left(1-\Delta\right)\right]
&\equiv&C_1\,=\,{\rm constant}
\label{textcontinuity}
\eqe
[see Eq.~(\ref{continuity5}), noting that in 
this section we use $\gamma$ and $v$ to refer 
to the zeroth order quantities, denoted in the Appendix by 
$\gamma_0$ and $v_0$]. 
\subsection{The energy equation}
After time averaging, the energy equation is 
\eqb
r^2\left<
w'\gamma^2v+\frac{EB}{4\pi}c\right>&=&{\rm constant}
\label{textenergya}
\eqe
Here $w'$ is the enthalpy density: outside of 
the sheets $w'=m\ncold c^2$, whereas within them $w'=4p'+m\nhot c^2$ 
($=B'^2/2\pi+m\nhot c^2$). One can substitute for the 
electric field using $E=vB$
because, by assumption, $E$ and $B$ are nonzero only outside the sheets,
where magnetic field is frozen into the plasma.

Replacing once again in Eq.~(\ref{textenergya}) 
the time average by a spatial one yields
\eqb
\gamma^2vr^2\left[\left(\frac{B'^2}{2\pi}
+{m\ncold c^2\over Z}\right)\Delta+(1-\Delta)
\left(m\ncold c^2+\frac{B'^2}{4\pi}\right)\right]&\equiv&C_2\,=\,
{\rm constant}
\label{textenergy}
\eqe
[see Eq.~(\ref{energy5})].

\citet{coroniti90} did not use equation (\ref{textenergya}) 
in his analysis.
Instead, he assumed the plasma both inside and outside 
of the sheet is strictly stationary 
in the wave frame so that the sheet edge moved through a constant density 
plasma. This led him to set $\nhot=\ncold$ in  
Eq.~(\ref{textcoroniticond}) (our numbering), 
thus reducing the number of unknowns. He was then able to 
find an expression for the speed of the sheet edge.
However, such a picture violates energy conservation, as is apparent from 
Eq.~(\ref{textenergya}): 
the condition of pressure 
equilibrium ($p'=B^2/8\pi\gamma^2$) means that a decrease in 
the enthalpy flux
in the magnetized part of the flow ($=r^2 v B^2/4\pi$) cannot 
be balanced by an increase in
the enthalpy flux in the sheet ($=r^2 w'\gamma^2 v=4r^2 p'\gamma^2v$)
unless there is a velocity (and density) jump across the sheet edge. 

\subsection{The entropy equation}
The entropy equation requires more care. Following Eq.~(21) of 
\citet{coroniti90} and setting the ratio of specific heats to $4/3$,
we write for the full nonlinear equation:
\eqb
3{\diff p'\over \diff t}-4{p'\over n'}
{\diff n'\over \diff t}&=&\left(E-{v\over c}B\right)j
\label{textentropy}
\eqe
(see Eq.~\ref{entropy})
where the convective derivative 
$d/dt\equiv\partial/\partial t + v\partial/\partial r$. 
The right-hand side of this equation, when multiplied by $\gamma$, is 
the rate of entropy generation in the rest-frame of the plasma, which moves
with speed $v$. 
In the limit of a sharp transition between the magnetized and 
unmagnetized parts of the flow, the entropy generation 
term is the product of two singular functions: 
one for the current and another for the electric field in the plasma
frame. To find the slow evolution of the wave, it is essential to 
perform the averaging process {\em before} inserting this specific
representation, which is not well-defined at the point
at which entropy is generated. In moving from his Eq.~(21) to (22), 
Coroniti 
overlooked this point. As a consequence of this, and of 
the incorrect expression for the expansion speed of the sheets, 
he came to the erroneous conclusion [in his equation (26)]
that the wave Lorentz factor remains almost constant during reconnection. 

Following the averaging procedure described in the Appendix 
(Eqs.~\ref{entropy3b} to \ref{entropy4b}),
we find
[see Eq.~(\ref{entropy5b})]:
\eqb
{4p'\over r^2}{\partial\over\partial r}
\left(r^2 \gamma v\Delta\right)
+3\gamma v{\partial\over\partial r}\left(\Delta p'\right)
+&&
\nonumber\\
{v\over\gamma}{\partial\over\partial r}
\left[\gamma^2\left(1-\Delta\right)p'\right]
+{2\gamma p' \left(1-\Delta\right)\over r}{\partial\over\partial r}
\left(r v\right)
&=&0\enspace.
\label{textentropy1}
\eqe
This equation may be integrated, 
using the equations of continuity (\ref{textcontinuity}) and flux freezing
(\ref{textfreeze}):
\eqb
(1-3Z)\ncold r^3&\equiv&C_3\,=\,{\rm constant}
\label{textentropy2}
\eqe
[See Eqs.~(\ref{entropy6}--\ref{thirdintegral})].

The system thus consists of five algebraic equations 
(\ref{textfreeze}),
(\ref{textcoroniticond}),
(\ref{textcontinuity}),
(\ref{textenergy}) and
(\ref{textentropy2}) for the five slowly varying 
unknown functions of 
radius:
$\ncold$,
$Z$,
$B'$,
$\Delta$ and
$\gamma$.

\section{Results}
\label{results}
\subsection{Asymptotic solution}
A general solution to this system is difficult to find, and the 
most straightforward way to generate a numerical solution is to revert to
integrating the differential forms of the continuity, energy and entropy 
equations. However, it is possible to extract analytically an asymptotic 
solution, valid for $\Delta$, $\gamma^{-2}$, $\slight^{-1}\ll 1$, which is just the 
regime we are interested in. 

As can be checked {\em a posteriori}, 
the quantity $\ncold r^3$ is an increasing 
function of $r$, so that, according to 
Eq.~(\ref{textentropy2}), $Z\rightarrow 1/3$ as $r\rightarrow\infty$.
Defining, in accordance with Eq.~(\ref{sigmal}),
\eqb
\slight&=&\left[{B'^2/4\pi\over m\ncold c^2}\right]_{r=\rlight}
\eqe
we can use the continuity equation (\ref{textcontinuity}) to
rewrite the energy equation (\ref{textenergy}) as
\eqb
\gamma^2 v r^2(1+\Delta){B'^2\over 4\pi} + \gamma m c^2 C_1&=&
\glight m c^2 C_1 (1+\slight)
\enspace.
\label{textenergy2}
\eqe
Here we have assumed the initial thickness of the current sheet is 
vanishingly small. This is reasonable, since before the onset of 
reconnection, the plasma in the sheet expands adiabatically and
cools, causing the sheet width to decrease. 
To lowest order in the small parameters, 
this equation (\ref{textenergy2}), together with Eq.~(\ref{textfreeze}), 
and the continuity equation (\ref{textcontinuity})
yield the same result:
\eqb
\gamma r^2 \ncold&=& C_1
\eqe 
so that the system is nearly degenerate.
Expanding the equations to next order and 
eliminating the leading term
gives
\eqb
\slight\left({2\Delta\over Z}-3\Delta+{1\over2\glight^2}
-{1\over2\gamma^2}\right)&=&{\gamma\over\glight}-1
\enspace.
\label{leadingtermout}
\eqe 
At large radius $Z\rightarrow 1/3$ and 
for a super-Alfv\'enic flow 
($\gamma\gg\glight>\sqrt{\sigma}$), one finds
$\gamma/\glight\rightarrow3\slight\Delta$, leading to the asymptotic solution 
\eqb
\gamma&=&\slight\glight\left[
{\Omega r\over 2\pi\olight\rlight}\right]^{1/2}
\label{gammaasympt}\\
\Delta&=&\left[{\Omega r\over 18\pi\olight\rlight}\right]^{1/2}
\label{deltaasympt}\\
{T'\over mc^2}&=&{\slight\glight\over6\gamma}\,=\,
\left[{18\Omega r\over \pi\olight\rlight}\right]^{-1/2}
\label{tempasympt}\\
p'&=&\left({\blight^2\over 8\pi\glight^2}\right)
\left({r\over\rlight}\right)^{-3}{2\pi\olight\over\Omega}
\label{pressasympt}
\eqe
which agrees with our estimate that the maximum value of $\gamma$ given 
in Eq.~(\ref{gammamax}) is attained at the radius $r_{\rm max}$ of  
Eq.~(\ref{correctrmax}), where $\Delta\sim1$. Note that this
solution is independent of the actual value of
$\glight$, which enters only as a scaling factor for $\gamma$. 

\subsection{Numerical solution}
\begin{figure}
\includegraphics[bb=30 157 580 706,width=\textwidth]{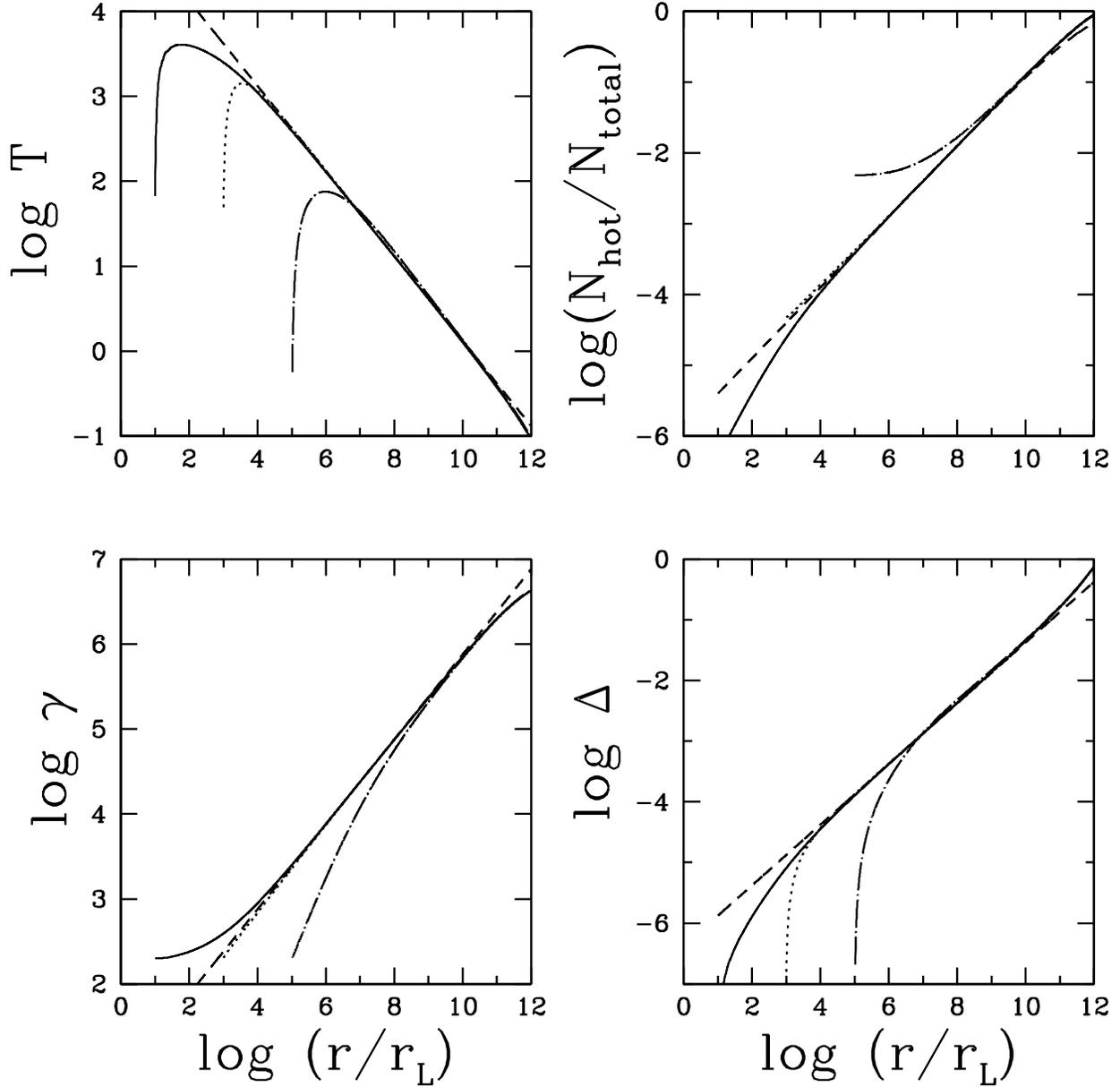}
\caption{\protect\label{numerical}
Results of a numerical integration 
the system of equations 
(\protect\ref{continuity5}), (\protect\ref{energy5}) and 
(\protect\ref{entropy5b})
for three different initial conditions, superposed on 
the analytic asymptotic solution (dashed line).
The parameters used are those appropriate for the Crab: $\slight=3\times10^4$,
$\glight=200$, $\olight/\Omega=10^{11}$. Plotted are the temperature 
$T$ in the hot sheet, in units of $mc^2$, the fraction $\alpha$ of particles in
the sheet [$\Delta\nhot/[\Delta\nhot+(1-\Delta)\ncold]$] the Lorentz factor of the flow 
$\gamma$ and the fraction 
of a wavelength occupied by the sheets 
$\Delta$, as functions of the 
radius in units of the light cylinder radius. The 
initial conditions correspond to reconnection starting at $r/\rlight=10$
 (solid line), $10^{3}$ (dotted line) and $10^{5}$ (dashed-dotted line).
In the Crab, the termination shock is located at $\log(r/\rlight)\approx 9$.
}
\end{figure}

Three physical parameters determine the cold radial MHD wind of a pulsar in the
absence of reconnection. They are the values at the light cylinder of 
the magnetization parameter 
$\slight$, Lorentz factor $\glight$ and the ratio of the particle
gyro frequency to the rotation period $\olight/\Omega$. These are
related to the multiplicity parameter $\kappa$ by 
Eq.~(\ref{sigmal}).
In the presence of reconnection, the same three parameters also uniquely 
specify the asymptotic solution at large radius
Eqs.~(\ref{gammaasympt}--\ref{pressasympt}). 
However, the full solution requires one additional initial condition, which is
the fraction $\alpha$ of plasma which is initially present in the 
current sheets. This quantity determines the radius at which 
reconnection starts, which is, formally, the position at which we impose
the initial condition $\Delta=0$. The perturbation 
analysis presented above is valid when this radius is large 
compared to $\rlight$.

To investigate the dependence of the solutions on this initial condition, 
we have solved the system by integrating numerically the three 
equations continuity, energy and entropy written in differential form: 
(\ref{continuity5}), (\ref{energy5}) and (\ref{entropy5b}). 
The results are shown in Fig.~\ref{numerical}, for parameters 
appropriate for the Crab pulsar: $\slight=3\times10^{4}$, 
$\glight=200$, ($>\sqrt{\slight}$, so that the wind is initially 
supersonic) and $\olight/\Omega=10^{11}$. The multiplicity 
parameter for these parameters is $\kappa=8\times10^3$.
Solutions for three different initial conditions are shown, corresponding to 
starting points for reconnection at $10$, $10^3$ and $10^5\times\rlight$. 
Superposed on these solutions is the analytic asymptotic solution. For all 
values of the initial condition, the asymptotic 
solution is approached rapidly, 
and followed closely, until $\Delta\sim1$. The termination shock in the 
wind of the Crab pulsar is thought to lie at roughly 
$r_{\rm s}\approx10^{9}\rlight$. At this point, $\Delta\sim 0.01$ and the 
energy flux is still dominated by Poynting flux. 

\subsection{Validity of the solutions}
The range of validity of our treatment is limited by two factors.
According to our solutions, the temperature in the current
sheets decreases outward. The expression used for the gyro-radius
in Eqs.~(\ref{textcoroniticond}) and (\ref{coroniticond})
assumes $T>mc^2$, as does the choice $\hat\gamma=4/3$ for the ratio of specific heats. 
This is not the case at very large radius
and the range of validity is therefore restricted to 
\eqb
\gamma&<&{\gamma_{\rm max}\over 6}
\eqe
At the upper end of this range, $\Delta=1/18$, 
so the assumption $\Delta\ll 1$
is still valid. The corresponding radius $\rnonrel$ is
\eqb
\rnonrel&=&
\frac{\pi\olight}{18\Omega}\rlight=\frac{10^6\mu_{30}}{P^2}\rlight
\eqe
At this point, the fraction of the Poynting flux transferred
to the plasma is still small. 
For the Crab, one obtains (see Fig.~\ref{numerical})
$\rnonrel\approx10^{10}\rlight$,
which already exceeds the radius of the standing shock. Thus,
our treatment is valid up to the shock front, 
before which 
reconnection is indeed ineffective in the 
wind of this pulsar.

The second limitation concerns the assumption that the 
dissipation proceeds sufficiently quickly to keep the 
sheet thickness equal to the minimum value given by
Eq.~(\ref{larmor}).
This assumption holds provided the proper propagation time, $r/(c\gamma)$,
exceeds the Larmor period. Because the sheet width is equal to the
Larmor radius, this condition,
\eqb
\frac r{\gamma}&>&4\pi^2\Delta\gamma\rlight
\label{crossing}
\eqe
is equivalent, to within a factor of the order of unity, to
the condition that pressure equilibrium within the sheet has
time to be established. 
Substituting Eq.~(\ref{deltaasympt}) and $Z=1/3$,
one finds 
\eqb
\gamma&<&\gamma_{\rm M}\,=\,3\kappa.
\eqe
If $\gamma_{\rm max}>\gamma_{\rm M}$, 
the flow accelerates only up to the point where $\gamma =\gamma_{\rm M}$,
after which no further dissipation can occur. 
This regime arises for
\eqb
\kappa&<&\sqrt{\olight\over6\Omega}
\\
&=&2\cdot10^3{\sqrt{\mu_{30}}\over P}
\eqe
or, alternatively,
\eqb
\gamma_{\rm max}&=&\slight\glight\,>\,\sqrt{3\olight\over2\Omega}
\\
&=&6\cdot10^3{\sqrt{\mu_{30}}\over P}
\eqe
Otherwise, acceleration continues and the flow can,
in principle, reach $\gamma_{\rm max}$, unless it encounters a shock front.

In the case of a pulsar moving 
through the interstellar medium, 
one can estimate the position of the termination shock by equating the 
magnetic pressure to the ram pressure of the medium:
\eqb
{\blight^2\over 8\pi}\left({\rlight\over r}\right)^2&=&\rho V^2
\eqe
where $\rho$ is the density of the 
interstellar medium, and $V$ the velocity of the pulsar through it. 
Taking $V=100\,{\rm km\,s}^{-1}$ and a particle density of 
$1\,{\rm cm}^{-3}$, one obtains
\eqb
r_{\rm s}&=&7\times 10^{14}{\mu_{30}\over P^2}\,{\rm cm}
\enspace.
\eqe
Comparing this with the expression for $r_{\rm max}$, 
Eq.~(\ref{correctrmax}), one sees that the flow remains Poynting 
dominated up to the termination 
shock for all but the milli-second pulsars.

\section{Conclusions}
\label{conclusions}

We have examined the fate of a wave generated in a pulsar wind
by the rotating, magnetized neutron star.
This wave is built from the oscillating equatorial
current sheet which, at large distances, may be considered
locally as a sequence of 
spherical current sheets separated in radius by the distance
$\pi\rlight$. 
As was suggested by \citet{usov75} and \citet{michel82},
the wave decays because the particle number density
eventually becomes insufficient to maintain the necessary current.
Dissipation begins when the velocity of the current carriers 
reaches the speed
of light or, almost equivalently, when the sheet width becomes
equal to the Larmor radius \citep{coroniti90, michel94}. The 
dissipation process may be considered as reconnection of 
oppositely directed magnetic fields \citep{coroniti90}. 
In the proper
plasma frame, plasma from the inter sheet space slowly moves towards
the sheet, which slowly expands, absorbing more plasma and
magnetic energy. 
The distance at which the wave decays completely 
is proportional to the Lorentz-factor $\gamma$ 
of the flow [Eq.(\ref{gammamax})], as found
by Michel and Coroniti. 
In this formula, however, they 
inserted  the initial Lorentz-factor, $\gamma\sim 100-1000$
and concluded that in the case of the Crab pulsar
the wave decays well before the wind reaches the termination shock.

We find that the flow accelerates during the dissipation
process. 
The reason is that in the freely expanding flow, hot plasma 
in the current sheet performs work on the flow. 
The restraining 
magnetic tension force is released by reconnection, but the accelerating 
pressure gradient remains. As a result, most of the magnetic energy is
dissipated when the flow has accelerated to a Lorentz-factor which is
of the same 
order as the maximal one.  For the Crab, the corresponding
distance is well beyond the standing shock, so we conclude that the
wave does not dissipate before entering the shock.
Using a simple estimate of the location of the termination shock
for other isolated pulsars, 
we find this conclusion holds 
for all except those of milli-second period. 
In the Appendix, we show that our result applies not only to the 
singular current sheet structure discussed 
by \citet{coroniti90}, but also to a more general 
smooth distribution of current and magnetic field.
 
Exactly how the wind energy dissipates remains a mystery, not only in the case of pulsars, 
but also in the closely related models proposed for gamma-ray bursts
\citep{usov92,usov94,blackmanyi98}.
At present, one 
can only 
speculate that dissipation might be possible in a combination of 
shocks and current sheets at the position of the 
bright equatorial X-ray torus observed in
the Crab \citep{brinkmannetal85}. 
If the Crab is an oblique rather than a perpendicular rotator,
a significant part of the energy flux at high latitudes is
transferred by the axisymmetric part of the Poynting flux, which cannot
dissipate by reconnection. It has been suggested that
the release of 
this energy may be triggered by the kink instability 
\citep{lyubarsky92, begelman98}, giving rise to 
the jet-like structure observed in the Crab Nebula, 
orientated, presumably, along the rotation axis of the 
pulsar \citep{hesteretal95}. Thus, the idea that dissipation 
of the axisymmetric component and the wave component 
of the Poynting flux
is fundamentally different has both observational 
and theoretical support. 
Current sheets are not expected to be responsible for the former. We
have demonstrated in this paper that they also cannot be responsible 
for the latter, until the pulsar wind 
encounters the termination shock.  

\acknowledgments
Y.L. thanks the Max-Planck-Institut f\"ur Kernphysik for support under 
their international visitor program.

\appendix
\section{Equations of entropy-wave evolution 
in the two-timescale approximation}
\subsection{MHD equations}
Consider a non-steady, axially symmetric, radial 
MHD wind. In spherical coordinates, the electromagnetic 
fields, fluid velocity and current density 
in the laboratory frame are 
$\vec{B}=B(r,t)\unitphi$, 
$\vec{E}=E(r,t)\unittheta$, 
$\vec{v}=v(r,t)\unitr$, 
$\vec{j}=j(r,t)\unittheta$, 
and the proper (i.e., in the fluid rest frame) energy density, pressure, 
temperature (in energy units) and 
number density are
$e'(r,t)$,
$p'(r,t)$
$T'(r,t)$ and 
$n'(r,t)$.

Then the equations of MHD are that of continuity:
\eqb
{\partial\over \partial t } (\gamma n')
+{1\over r^2}{\partial\over \partial r }(r^2\gamma vn')
&=&0
\label{continuity}
\eqe
(where $\gamma=1/\sqrt{1-v^2/c^2}$),
the energy equation (zeroth component of the 
divergence of the stress-energy tensor):
\eqb
{\partial\over \partial t } (T^{00})
+{1\over r^2}{\partial\over \partial r }(r^2T^{01})
&=&0
\label{energy}
\eqe
where 
\eqb
T^{00}&=&(e'+p')\gamma^2 - p' + {E^2+B^2\over 8\pi}
\nonumber\\
T^{01}&=&(e'+p')\gamma^2v + {EBc\over 4\pi}
\label{defstress}
\eqe
the entropy equation:
\eqb
{1\over\hat\gamma-1}\left(
{\diff p'\over \diff t}-{\hat\gamma p'\over n'}{\diff n'\over\diff t}
\right)&=&\left(E-{v\over c}B\right)j
\label{entropy}
\eqe
(where $\diff /\diff t\equiv\partial/\partial t + v\partial/\partial r$
and $\hat\gamma$ is the ratio of specific heats)
and the two Maxwell equations 
\eqb
{1\over r}{\partial\over \partial r }(rB) 
+ {1\over c}{\partial E\over \partial t }
+{4\pi\over c}j
&=&0
\label{ampere}
\eqe
(Amp\`ere's law)
and
\eqb
{1\over r}{\partial\over \partial r }(rE) 
+ {1\over c}{\partial B\over \partial t }
&=&0
\label{faraday}
\eqe
(Faraday's law).
The system is completed by Ohm's law
\eqb
j&=&\conduct\gamma\left(E-{v\over c}B\right)
\label{ohm}
\eqe
the ideal gas law 
\eqb
p'&=&n'T'
\label{idealgas}
\eqe
and an equation of state, for which we select
\eqb
p'&=&(\hat\gamma -1)(e'-n' mc^2)
\label{eqofstate}
\eqe
where $m$ is the (mean) particle rest mass. In the following
we will take $\hat\gamma=4/3$, as appropriate for a relativistic gas.

Eliminating $n'$ 
from Eq.~(\ref{entropy}) 
using the equation of 
continuity (\ref{continuity}) we find
\eqb
4p'{\partial\gamma\over \partial t}+
3\gamma{\partial p'\over \partial t}+
{4p'\over r^2}{\partial\over \partial r}(r^2\gamma v)+
3v\gamma{\partial p'\over\partial r}&=&\gamma\left(E-{v\over c}B\right)j
\enspace.
\label{newentropy}
\eqe
In the limit of a relativistic gas, $e'\rightarrow3p'$,  the set of 
equations (\ref{energy}), 
(\ref{newentropy}),
(\ref{ampere})
and
(\ref{faraday}) is then independent of $n'$, which appears only in the 
continuity equation (\ref{continuity}).
 
\subsection{Perturbative solution}
To solve these equations, 
we exploit the two-timescales present in the problem. The fast timescale
is that of the pulsar rotation period $t_{\rm fast}=P\equiv2\pi/\Omega$. 
We assume that the initial conditions 
in the wind close to the star have period $P$ and 
that at any fixed radius, all quantities 
vary with this period. 
At a distance of 
a few light-cylinder radii ($\rlight\equiv c/\Omega$), 
we assume that 
the wind has settled down into an almost stationary pattern in which 
the fluid speed does not vary on the fast timescale, although the density, 
pressure and especially the magnetic field do so.
Our initial conditions at this distance, therefore, 
constitute an entropy wave comoving 
with the fluid. The wave evolves on the 
slow timescale which is the expansion timescale of the wind
$t_{\rm slow}\sim r P/\rlight$. For $r\gg \rlight$, we have conditions 
suitable for a two-timescale expansion, i.e., $t_{\rm fast}\ll t_{\rm slow}$.
The general procedure is: i) transform to fast and slow independent variables,
ii) expand the dependent variables, iii) collect and solve the zeroth order
equations, and iv) collect the first order equations and demand that
the secular terms they contain vanish \citep{nayfeh73}.

First, we define a fast phase variable
\eqb
\phi&=&\Omega\left[t- \int_0^r{\diff r'\over \vwave(r')}\right]
\eqe
where $\vwave(r)$ is the speed of the pattern, which is to be determined.
We now change variables from 
$(r,t)$ to $(\phi,\rslow)$, where the dimensionless
\lq slow\rq\ variable is defined as
\eqb
\rslow&=&\eps r/\rlight
\eqe
with $\eps\ll 1$ and $\rslow\sim1$.
In terms of the new variables, we find for the 
continuity equation~(\ref{continuity}):
\eqb
{\partial\over\partial\phi}(\gamma n') -
{1\over \vwave}{\partial\over\partial\phi}(\gamma vn') +
{\eps\over\rslow^2}{\partial\over\partial \rslow}(\rslow^2\gamma v n')
&=&0
\label{continuity1}
\eqe
the energy equation~(\ref{energy}):
\eqb
{\partial T^{00}\over\partial\phi} -
{1\over \vwave}{\partial T^{01}\over\partial\phi} +
{\eps\over \rslow^2}{\partial\over\partial \rslow}(\rslow^2 T^{01})
&=&0
\label{energy1}
\eqe
the modified entropy equation~(\ref{newentropy}):
\eqb
\lefteqn{
4p'{\partial\gamma\over\partial\phi} +
3\gamma{\partial p'\over\partial\phi} -
4{p'\over \vwave}{\partial\over\partial\phi}(\gamma v) -}
\nonumber\\
&&
3{\gamma v\over\vwave}{\partial p'\over\partial\phi} +
{4\eps p'\over\rslow^2}{\partial\over\partial \rslow}(\rslow^2 \gamma v) +
3\eps \gamma v {\partial p'\over\partial \rslow}
\,=\,{1\over\Omega}\gamma(E-vB)j
\label{entropy1}
\eqe
Amp\`ere's equation~(\ref{ampere}):
\eqb
{\partial E\over\partial\phi} -
{1\over \vwave}{\partial B\over\partial\phi} +
{\eps\over\rslow}{\partial\over\partial \rslow}(\rslow B)+
{4\pi\over \Omega}j 
&=&0
\label{ampere1}
\eqe
Faraday's equation~(\ref{faraday}):
\eqb
{\partial B\over\partial\phi} -
{1\over \vwave}{\partial E\over\partial\phi} +
{\eps\over\rslow}{\partial\over\partial \rslow}(\rslow E) 
&=&0
\enspace.
\label{faraday1}
\eqe
In these equations we have 
expressed all velocities in units of the speed of light 
($v\rightarrow cv$ etc.).

We now expand the dependent variables, 
noting that for an entropy wave the zeroth order velocity
is independent of the phase:
\eqb
v&=&v_0(\rslow)+\eps v_1(\phi,\rslow)
\nonumber\\
n'&=&n'_0(\phi,\rslow)+\eps n'_1(\phi,\rslow)
\nonumber\\
p'&=&p'_0(\phi,\rslow)+\eps p'_1(\phi,\rslow)
\eqe
etc.

Substituting and collecting terms of order $\eps^0$ we find 
from the continuity equation (\ref{continuity1})
that for 
a non-uniform wind (one in which $\gamma_0n_0$ is a function of $\phi$)
\eqb
\vwave&=&v_0(\rslow)
\enspace.
\label{equalspeeds}
\eqe
The Maxwell equations (\ref{faraday1}) and (\ref{ampere1})
then lead to 
\eqb
{\partial B_0\over\partial\phi}&=&
{4\pi\gamma_0^2v_0\over\Omega}j_0
\enspace.
\label{ampere2}
\eqe
From Faraday's law (\ref{faraday1}), it then follows that the quantity 
$E_0-v_0B_0$ is independent of $\phi$. Furthermore, in order to describe a
wave with both a reconnection zone and a region in which ideal MHD holds
($\conduct\rightarrow\infty$), it follows from Ohm's law Eq.~(\ref{ohm}) that 
this quantity must be zero:   
\eqb
E_0&=&v_0B_0
\label{idealmhd}
\enspace.
\eqe
The energy equation (\ref{energy1})
yields the pressure balance condition:
\eqb
{\partial\over\partial\phi}\left(
p'_0+{B_0^2\over8\pi\gamma_0^2}\right)&=&0
\enspace.
\label{energy2}
\eqe
Finally, using Eqs.~(\ref{equalspeeds}) and (\ref{idealmhd}) 
it can be seen that 
the entropy equation (\ref{entropy1}) is satisfied identically 
to zeroth order.

To first order in $\eps$, we have for the continuity, energy and 
Maxwell equations:
\eqb
{\gamma_0\over v_0}{\partial \over\partial\phi}(n'_0v_1)&=&{1\over\rslow^2}
{\partial \over\partial\rslow}(\rslow^2\gamma_0v_0n'_0)
\label{continuity3}
\\
{\partial \over\partial\phi}\left({1\over v_0}T_1^{01}-T_1^{00}\right)
&=&
{\partial \over\partial\phi}\left(p'_1+{v_1\gamma_0^2\over v_0}(e'_0+p'_0)
+{B_0\over 4\pi\gamma_0^2v_0}E_1\right)
\nonumber\\
&=&
{1\over\rslow^2}{\partial \over\partial\rslow}\left(\rslow^2 T_0^{01}\right)
\label{energy3}
\\
{\partial \over\partial\phi}\left(E_1-{1\over v_0}B_1\right)&=&
-{1\over\rslow}{\partial \over\partial\rslow}(\rslow B_0)-{4\pi\over\Omega}j_1
\label{ampere3}
\\
{\partial \over\partial\phi}\left(B_1-{1\over v_0}E_1\right)&=&
-{1\over\rslow}{\partial \over\partial\rslow}(\rslow v_0B_0)
\enspace.
\label{faraday3}
\eqe
In the entropy equation, one may substitute for the 
zeroth order current using equation~(\ref{ampere2}) to find
\eqb
\lefteqn{
{4\gamma_0\over v_0}{\partial \over\partial\phi}(p'_0v_1)
+}
\nonumber\\
&&{1\over4\pi\gamma_0v_0}(E_1-v_0B_1){\partial B_0\over\partial\phi}\,=\,
{4p'_0\over\rslow^2}{\partial \over\partial\rslow}(\rslow^2\gamma_0v_0)+
3\gamma_0v_0{\partial p'_0 \over\partial\rslow}
\enspace.
\label{entropy3b}
\eqe

The equations governing the evolution of the zeroth order quantities on 
the slow scale are given in the usual manner
by imposing non-secular behavior on 
the first order equations \citep{nayfeh73}. 
This ensures that the first order 
quantities do not grow to dominate the zeroth order terms of the expansion 
within 
${\rm O}(r/\rlight)$ wave periods. 
As frequently happens, the imposition of these 
regularity conditions suffices to determine the slow 
variation of the zeroth order quantities.
Consider, for example, Eq.~(\ref{continuity3}), which is a linear, 
inhomogeneous equation for $v_1$. The right-hand side is, by construction,
a periodic function of $\phi$. Therefore, $v_1$ will grow with $\phi$ unless
the integral of the right-hand side over a complete period vanishes.
Applying these considerations also to Eq.~(\ref{energy3}) leads to the 
conditions:
\eqb
{\partial\over\partial\rslow}
\left(\rslow^2\gamma_0v_0\int_0^{2\pi}\diff\phi n'_0\right)&=&0
\label{continuity4}\\
{\partial\over\partial\rslow}
\left(\rslow^2 \int_0^{2\pi}\diff\phi T_0^{01}\right)&=&
{\partial\over\partial\rslow}
\left\lbrace\rslow^2\gamma_0^2v_0 
\left[\int_0^{2\pi}\diff\phi\left(e'_0+p'_0+{B_0^2\over 4\pi\gamma_0^2}
\right)\right]\right\rbrace
\nonumber\\
&=&0
\label{energy4}
\eqe
Equation (\ref{ampere3}) is needed only if the first order current
is required, and Eq.~(\ref{faraday3}) yields only the conservation of the 
phase-integrated flux.
Equation
(\ref{entropy3b}) is integrated by parts to give, 
using (\ref{faraday3})
\eqb
\lefteqn{
\int_0^{2\pi}\diff\phi p'_0
{4\over \rslow^2}{\partial\over\partial\rslow}
\left(\rslow^2 \gamma_0v_0\right)
+}
\nonumber\\
&&3\gamma_0v_0{\partial\over\partial\rslow}\int_0^{2\pi}\diff\phi p'_0
\,=\,-{1\over4\pi\gamma_0\rslow}\int_0^{2\pi}
\diff\phi B_0{\partial\over\partial\rslow}
\left(\rslow v_0 B_0\right)
\enspace.
\label{entropy4b}
\eqe

\subsection{Application to the striped wind}

For the striped wind (Fig.~\ref{sketch}), the zeroth order solution is
\[
\begin{array}{l}
      \left.\begin{array}{r@{\,=\,}l}
           n'_0 & \nhot(\rslow)\nonumber\\
           p'_0 & p'(\rslow)\nonumber\\
           B_0 & 0
       \end{array}\right\rbrace\quad
       {\rm for}\ 
0<\phi<\pi\Delta(\rslow)\ {\rm and\ } 
\pi<\phi<\pi[1+\Delta(\rslow)]
 \nonumber\\
      \left.\begin{array}{r@{\,=\,}l}
           n'_0 & \ncold(\rslow)\nonumber\\
           p'_0 & 0\nonumber\\
           B_0 & B(\rslow)
       \end{array}\right\rbrace\quad
       {\rm for\ }\pi\Delta<\phi<\pi\nonumber\\
 \nonumber\\
      \left.\begin{array}{r@{\,=\,}l}
           n'_0 & \ncold(\rslow)\nonumber\\
           p'_0 & 0\nonumber\\
           B_0 & -B(\rslow)
       \end{array}\right\rbrace\quad
       {\rm for\ }\pi[1+\Delta(\rslow)]<\phi<2\pi
\end{array}
\]
together with the condition of pressure equilibrium between the hot and cold
layers:
\eqb
p'&=&{B^2\over 8\pi\gamma_0^2}\,=\,{B'^2\over8\pi}
\label{pressureeqcond}
\eqe

Coroniti's estimate of the thickness of the neutral sheet gives
\eqb
\Delta(\rslow)&=&{p'(\rslow)\Omega\over \nhot(\rslow) e \pi B(\rslow) v_0(\rslow)}
\label{coroniticond} 
\eqe
and the ideal MHD condition outside the sheet implies 
flux freezing there:
\eqb
{\partial\over\partial \rslow}\left[{B(\rslow)\over \rslow\gamma_0(\rslow)\ncold(\rslow)}\right]&=&0
\enspace.
\label{freeze}
\eqe
This latter relation follows formally by using the first order form of the ideal MHD 
condition: $E_1=v_0B_1+v_1B_0$ in Faraday's equation~(\ref{faraday3}) 
together with the continuity equation~(\ref{continuity3}) and the fact that in
our present configuration $B_0$ and $n_0$ are independent of 
$\phi$ outside of the sheet.  

The five unknowns obey, in addition, equations 
(\ref{continuity4}, \ref{energy4}, 
and \ref{entropy4b}).
In the case of Eqs.~(\ref{continuity4}) and (\ref{energy4}) 
the integration over $\phi$ may be performed immediately to give:
\eqb
{\partial\over\partial\rslow}
\left\lbrace
   \rslow^2\gamma_0v_0
   \left[
      \left(
      1-\Delta
      \right)
   \ncold+\Delta\nhot
   \right]
\right\rbrace&=&0
\label{continuity5}\\
{\partial\over\partial\rslow}
\left\lbrace
\rslow^2\gamma_0^2v_0 mc^2
\left[\left(1-\Delta\right)\ncold+\Delta\nhot\right]
+2\rslow^2\gamma_0^2v_0\left(1+\Delta\right){B'^2\over8\pi} \right\rbrace&=&0
\label{energy5}
\eqe
[Eq.~(\ref{textcontinuity} and \ref{textenergy}].
The left-hand side of Eq.~(\ref{entropy4b}) is also straightforwardly 
integrated. On the right-hand side, however, we first rewrite the 
integration in terms of $B_0^2$:
\eqb
{4\Delta p'_0\over\rslow^2}{\partial\over\partial\rslow}\left(
\rslow^2\gamma_0 v_0\right) + 3\gamma_0 v_0{\partial \over\partial\rslow}
\left(\Delta p'_0\right)&=&
-{1\over4\pi\gamma_0\rslow}\int_0^{2\pi}\diff\phi\left[
{\rslow v_0\over 2}{\partial B_0^2\over\partial\rslow} + 
B_0^2{\partial\over\partial\rslow}\left(\rslow v_0\right)\right]
\enspace.
\eqe
Now the integration over $\phi$ can be performed unambiguously to give,
using Eq.~(\ref{pressureeqcond}):
\eqb
{4\Delta B'^2\over \rslow^2}{\partial\over\partial\rslow}
\left(\rslow^2 \gamma_0v_0\right)
+3\gamma_0v_0{\partial\over\partial\rslow}\left(\Delta B'^2\right)
+&&
\nonumber\\
{v_0\over\gamma_0}{\partial\over\partial\rslow}
\left[\gamma_0^2\left(1-\Delta\right)B'^2\right]
+{2\gamma_0 B'^2 
\left(1-\Delta\right)\over\rslow}{\partial\over\partial\rslow}
\left(\rslow v_0\right)
&=&0
\enspace.
\label{entropy5b}
\eqe

To integrate the entropy equation (\ref{entropy5b}) we first  
multiplying it by $\rslow^2\gamma_0 v_0$ and combine terms
with and without $\Delta$ to find
\eqb
4\gamma_0 v_0{\diff\over\diff\rslow}
\left(\rslow^2v_0\gamma_0\Delta B'^2\right)-
2\gamma_0 v\rslow{\diff\over\diff\rslow}
\left(\rslow v_0\gamma_0\Delta B'^2\right)
+{\diff\over\diff\rslow}
\left(\rslow^2v_0^2\gamma_0^2 B'^2\right)&=&0.
\label{entropy6}
\eqe
The last term appears to be of zeroth order in $\Delta$, but, because of 
the continuity equation (\ref{continuity5}), it is in fact of first order. 
Substituting for $B'$ using the flux freezing 
condition of Eq.~(\ref{freeze}), this term can be reduced to
\eqb
2\rslow^2v_0\gamma_0 \ncold{\diff\over\diff\rslow}
\left(\rslow^2v_0\gamma_0 \ncold\right)&=&
2\rslow^2v_0\gamma_0 \ncold{\diff\over\diff\rslow}
\left(\rslow^2v_0\gamma_0\Delta (\ncold-\nhot)\right)
\eqe
where the continuity equation (\ref{continuity5}) was used. 
Returning to equation (\ref{entropy6}), and 
using the substitution
\eqb
v_0\gamma_0\Delta&=&{ \rslow Z\over 4\pi\kappa}
\eqe
where $Z$ is defined in Eq.~(\ref{defZ}),
one finds
\eqb
2\rslow\ncold{\diff Z\over \diff\rslow}
+3\ncold(3Z-1)+(3Z-1)\rslow{\diff\ncold\over\diff\rslow}=0.
\eqe
which integrates to
\eqb
(1-3Z)^{2/3}\rslow^3\ncold&=&{\rm constant}
\enspace.
\label{thirdintegral}
\eqe

The striped wind shown in Fig.~\ref{sketch} 
is a particular (and singular) idealization of a wind containing
cold magnetized parts of opposite polarity separated by hot sheets. More
generally, we can envisage an idealization in which $n'_0$, $B_0$ and
$p'_0$ are all continuous functions of $\phi$. 
First, assume $B_0$ and $n'_0$ take on the 
constant values $B$ and $\ncold$ 
outside of the sheets. From the condition of  pressure balance
\eqb 
B^2&=&B_0^2(\phi)+8\pi\gamma_0^2 p'_0(\phi)
\eqe
Defining the effective sheet width as
\eqb
\Delta&=&{8\pi\gamma_0^2\over B^2}\int_0^{2\pi}
{\diff \phi\over2\pi} p'_0(\phi)
\eqe
and the average particle density $\nhot$ within the sheet via:
\eqb
\left(\nhot-\ncold\right)\Delta&=&\int_0^{2\pi}
{\diff \phi\over2\pi} \left[n'_0(\phi)-\ncold\right]
\eqe
we find, from Eqs.~(\ref{continuity4}--\ref{entropy4b}) a system of 
equations which is identical to those obeyed in the singular idealization.

\end{document}